\documentclass[a4paper,11pt]{article}
\pdfoutput=1 

\usepackage{jcappub} 
\usepackage{bigints}
\usepackage[T1]{fontenc} 
\usepackage[normalem]{ulem}
\usepackage{color}

\title{\boldmath Constraints on polynomial inflation under power-law perturbations}

\author[a]{Maria E. S. Antunes,}
\author[b,c]{Micol Benetti,}
\author[a,1]{Eduardo Bittencourt,\note{Corresponding author.}}
\author[a]{Fernando A. Franco}

\affiliation[a]{Federal University of Itajub\' a, BPS Av. 1303, Itajub\'a, MG, Brazil}
\affiliation[b]{Scuola Superiore Meridionale, Largo S. Marcellino 10, I-80138 Napoli, Italy}
\affiliation[c]{Istituto Nazionale di Fisica Nucleare (INFN), sez. di Napoli, Via Cinthia 9, I-80126 Napoli,
Italy}

\emailAdd{eduardaantunes.012@gmail.com }
\emailAdd{m.benetti@ssmeridionale.it}
\emailAdd{bittencourt@unifei.edu.br}
\emailAdd{fernandoaf@unifei.edu.br }

\abstract{We investigate perturbations in power-law monomial potentials within inflationary models driven by a single scalar field. By introducing a second term to the original potential, we study how this perturbation influences the slow-roll parameters and analyze the consequent changes in the spectral index, $n_s$, and the tensor-to-scalar ratio, $r$, treating the additional term as a correction to the monomial case. Comparing our numerical results with current cosmological data from the Planck satellite observations on $n_s$, $r$, and the clustering parameter, $\sigma_8$, we place significant constraints on the free parameter of our class of inflationary potentials. We found that the perturbative consistency method we analyze could be an interesting test to explore for more complex inflationary models, looking for features that better match the data that might highlight fundamental physics implications.}

\begin{document}

\maketitle
\flushbottom

\section{Introduction}\label{sec:intro}
While highly successful, the $\Lambda$CDM model exhibits some shortcomings in the early era, particularly concerning the fine-tuning of its free parameters to explain the homogeneous and isotropic distribution of matter at large scales today \cite{Dodelson_2003,Mukhanov_2005}. To overcome these challenges, the inflationary mechanism, a short period of accelerated expansion proposed by Starobinsky, Guth, and others in the early Eighties \cite{Starobinsky80,Kazanas1980,Sato1981PLB,Sato1981MNRAS,Guth81,Albrecht1982,Linde1982}, became a widespread solution. The simplest way to implement this idea is through scalar fields driven by a given potential. This marked a turning point in cosmology, as it provided particle physicists with an appropriate scenario to test the consistency of their models before modern particle accelerators were developed. Since then, numerous inflationary models have been proposed in the literature, for instance, non-minimal couplings \cite{Lucchin1986,Salopek1989,Fakir1990}, string theory \cite{Garcia2002,Kachru2003,Baumann2009,Santos:2022exm,SantosdaCosta:2020dyl}, extra degrees of freedom \cite{Gil2016,Berera2023,Kamali2023,Rodrigues:2023kiz,Landau:2022mhm,Benetti:2019kgw,Campista:2017ovq,Benetti:2016jhf}, among others (see reviews \cite{Martin_2014} and \cite{Tsujikawa2014} for more details).

With the advent of precision cosmology, i.e the WMAP \cite{Hinshaw_2013,Bennett_2013} and Planck \cite{Planck2018_const_infl,Planck2018_grav_lens,Planck2018_VI} collaborations over the years, several inflationary models started being ruled out by the data. The critical point is that the slow-roll regime of inflation narrows the window for the cosmological parameters, making it arduous to adjust the free parameters of the theories with the observations. Although it seems that data favor the simplest class of inflationary models driven by a single field, the most viable ones are given by highly nonlinear phenomenological potentials \cite{Planck2018_const_infl}. In particular, the Starobinsky inflation \cite{Starobinsky80} is still the best-suited model\footnote{From the historical point of view, the first model \cite{Novello79} able to circumvent the problems of the early universe gave birth to the \textit{bouncing cosmology}. Nowadays, cosmologists try to push this forward as an alternative to the standard cosmological scenario (see details in reviews \cite{Novello2008,Peter2017}).}.

In this paper, we investigate models of the single scalar field where the Lagrangian has a canonical kinetic term and a polynomial potential representing a truncated Taylor expansion. By reproducing the well-established results of the power-law potentials initially, we thus introduce a second term to the monomial potential and study the variations of the slow-roll parameters due to such a perturbation. Finally, we determine which kind of term is preferred by the data according to a given leading term in the potential. In other words, as it is expected that any potential for the scalar field is an analytic function admitting a power series expansion, we study the sensitivity of the spectral index, $n_s$, and the scalar-to-tensor ratio, $r$, in the case of binomial potentials where the second term is seen as a correction to the first one. This kind of analysis deserves special attention, as it may, on the one hand, aid in the search for other physically meaningful potentials allowed by Planck data, and on the other hand, indicate potentials that emerge from an effective field theory limit in a more fundamental context.

The paper is summarized as follows: in section \ref{sec:review_infl}, we introduce the mathematical notation used to construct the inflation with single scalar fields. Then, we briefly review the main results of slow-roll parameters when the scalar field has a monomial potential in sec. \ref{sec:monomial}. In section \ref{sec:binomial} we study two cases of binomial potentials according to the function parity of the terms involved. Finally, we present the concluding remarks in section \ref{sec:results}. Throughout the text, we set $\hbar=c=1$.

\section{Brief review of inflationary physics with a single scalar field}\label{sec:review_infl}

Let us begin by reviewing the basic equations that govern primordial inflation, which we will later apply to a simple monomial potential and to the more complex case of a binomial potential.

During the inflationary era, we assume as source a minimally coupled scalar field $\phi$ such that total action is taken as
\begin{equation}
\label{eq:action}
S = \int\sqrt{-g}\,d^4x\left(\frac{1}{2}M_{\textrm{pl}}^2\,\mathcal{R}-\frac{1}{2} g^{\mu\nu}\nabla_{\mu}\phi\nabla_{\nu}\phi- V(\phi)\right),
\end{equation}
where $\mathcal{R}$ is the Ricci scalar, $V(\phi)$ is the potential encoding of the possible self-interacting terms of the scalar field, and $M_{\textrm{pl}}$ is the Planck mass.

The time evolution of the universe is driven by the Friedmann-Lema\^itre-Robertson-Walker (FLRW) background metric
\begin{equation}
\label{eq:friedmann}
ds^2 = -dt^2 + \dfrac{a^2(t)}{1-kr^2}dr^2 + a^2(t)r^2(d\theta^2 + \sin^2\theta d\varphi^2),
\end{equation}
where $a(t)$ is the scale factor and $k$ is the constant spatial curvature. Thus, the equations of motion for $\phi$ and $a(t)$ are straightforward from variational methods applied to the equation \eqref{eq:action}. In particular, variations of $S$ with respect to $\phi$ yields
\begin{equation}
\label{eq:ce}
\ddot\phi + 3H\dot\phi+V_{,\phi} = 0,
\end{equation}
where $\phi$ is a function of time and the dot means derivative with respect to the time coordinate.

Variations of $S$ with respect to the metric $g_{ab}$ gives the Einstein field equations $G_{ab}=(1/M_{\textrm{pl}}^2)\,T_{ab}$, from which the nontrivial components are
\begin{equation}
\label{eq:EFE1}
H^2 = \dfrac{\rho}{3M_{\textrm{pl}}^2}-\dfrac{k}{a^2},
\end{equation}
and
\begin{equation}
\label{eq:EFE2}
\dot H = -\dfrac{1}{2M_{\textrm{pl}}^2}(\rho + p)+\dfrac{k}{a^2},
\end{equation}
where $H(t)=\dot a/a$ is the Hubble parameter, $\rho$ is the energy density and $p$ is the pressure, both determined in terms of a class of normalized, time-like observers of the form $V^{a}=\delta^{a}_0$. Their explicit expressions are, respectively,
\begin{equation}
\label{eq:rho_p_ce}
\rho = \frac{1}{2}\dot\phi^2 +V,\quad \mbox{and}\quad p = \frac{1}{2}\dot\phi^2-V.
\end{equation}

Regardless of the specific model, the inflation mechanism requires a slow-roll regime at the initial conditions to sufficiently increase the causal regions and explain the current homogeneity and isotropy of Cosmic Microwave Background (CMB). The most common definitions of the first and second-order Hubble slow-roll parameters are
\begin{equation}
\label{slow_roll}
\epsilon = \frac{d}{dt}\left(\frac{1}{H}\right),\quad \mbox{and}\quad
    \eta = \frac{\dot \epsilon }{H\epsilon}
\end{equation}
Concerning a single scale field, it is required $\dot{\phi}^2 \ll |V|$ and $\ddot{\phi} \ll 1$. Therefore, these parameters can be translated into ones in terms of derivatives of the scalar field, as follows
\begin{equation}
\label{eq:epsilon_v}
    \epsilon_V = \frac{M_{\textrm{pl}}^2}{2} \left(\frac{V_\phi}{V}\right)^2,
\end{equation}
and
\begin{equation}
\label{eq:eta_v}
\eta_V = M_{\textrm{pl}}^2\,\frac{V_{\phi\phi}}{V}.
\end{equation}
During the inflationary era, these parameters are assumed to be small while the end of inflation is characterized by $\epsilon_V\approx 1$ and $|\eta_V|\approx1$. The value of $\phi$ when this happens is denoted by $\phi_e$.

Also, in terms of $\phi$ the number of e-folds can be written down as
\begin{equation}
\label{n_efolds}
 N =  \int_a^{a_e} \frac{da'}{a'} = \frac{1}{M_{\textrm{pl}}}\int_{\phi_e}^{\phi} \frac{d \phi}{\sqrt{2 \epsilon}}.
\end{equation}
For fixed $\phi$, this quantity gives the amount of increasing of the Universe along the inflationary period. In particular, there are two values of interest: the total amount of e-folds, denoted by $N_i$, which is associated with the value of the scalar field at the beginning of inflation $\phi_i$; and the other, denoted by $N_*$, corresponding to the moment in which a chosen pivot wave-number $k_*$ crosses the Hubble horizon, with the scalar field value denoted by $\phi_{*}$. Throughout this paper, we shall fix $N_*=55$ for a pivot scale of $0.05\,\mbox{Mpc}^{-1}$, the same fiducial values used by Planck collaboration \cite{Planck2018_const_infl}.

To compare with the CMB data, it is more useful to write two other parameters in terms of the slow-roll ones, namely, the spectral index
\begin{equation}
\label{eq:spectral_index}
    n_s = 1 - 6 \epsilon_V + 2 \eta_V
\end{equation}
and the scalar-to-tensor ratio
\begin{equation}
\label{eq:r}
    r = 16 \epsilon_V .
\end{equation}
Finally, the amplitude of the dimensionless power spectrum when the perturbation mode crosses the horizon is \cite{Baumann_McAllister_2015}
\begin{equation}
\label{eq:P_r_phi}
    \Delta^2_R(\phi)= \frac{V(\phi)}{24 \pi^2 \epsilon M_{pl}^4}\bigg|_{\phi=\phi_*}.
\end{equation}
We now apply this to some polynomial potentials, namely, monomial and binomial.

\section{The simpler case: Monomial inflation}
\label{sec:monomial}

\begin{figure}[ht]
    \centering
    \includegraphics[scale=0.45]{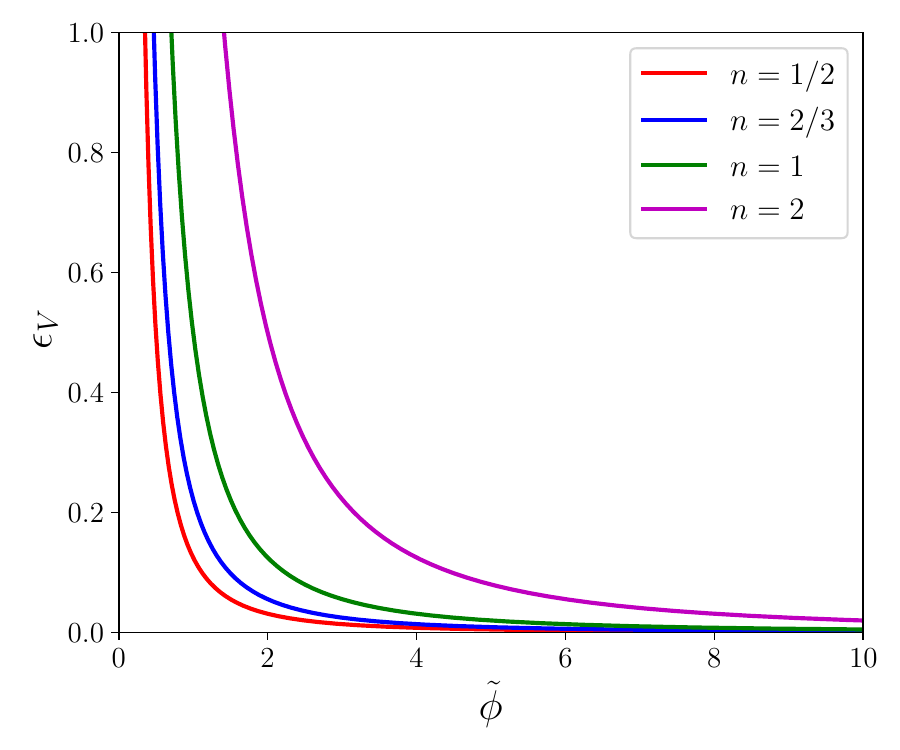}
    \includegraphics[scale=0.45]{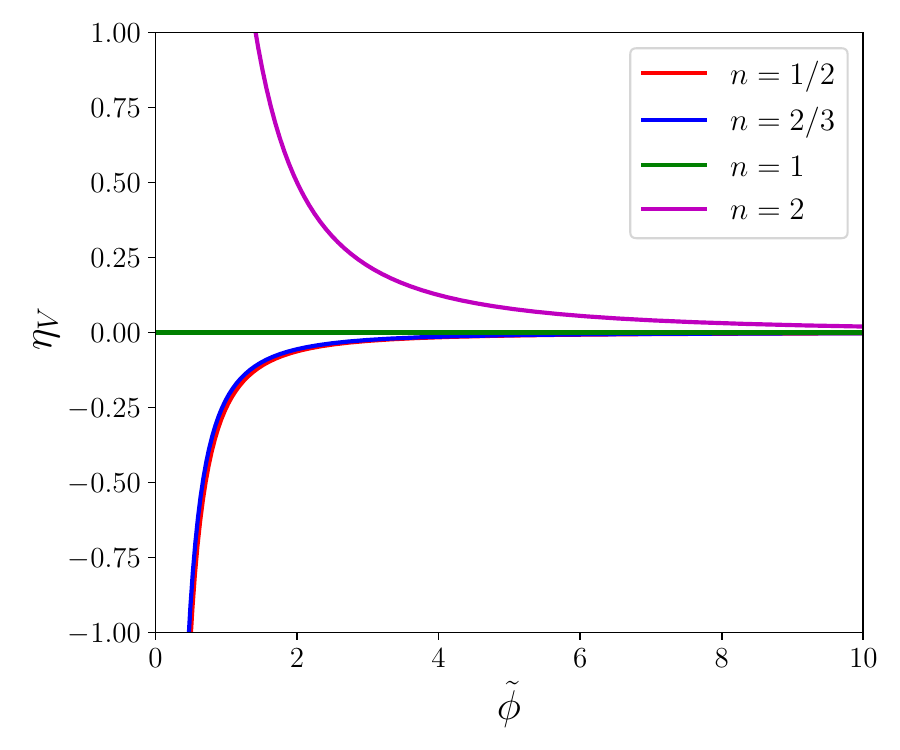}
    \caption{Slow roll parameters $\epsilon_V$ and $\eta_V$ as function of $\tilde\phi$ for a monomial potential.}
    \label{fig:slow_roll_monomial}
\end{figure}

We can now explore the straightforward case of a monomial potential, which will serve as the base for our approach. It is given by
\begin{equation}
\label{eq:mono_potential}
    V (\phi) = \alpha\, M_{{\rm pl}}^4\,\tilde\phi^n,
\end{equation}
where $\alpha$ is the amplitude of the potential, $n$ a positive real number, and we introduce $\tilde\phi=\phi/M_{{\rm pl}}$, for the sake of brevity. The slow roll parameters of the equations \eqref{eq:epsilon_v} - \eqref{eq:eta_v} are:
\begin{equation}
\label{eq:slow_roll_mono}
\epsilon_V = \frac{n^2}{2\tilde\phi^2},\quad \mbox{and} \quad
    \eta_V = \frac{n(n-1)}{\tilde\phi^2},
\end{equation}
with the invariance of the slow-roll parameters under the transformation $\tilde\phi\rightarrow -\tilde\phi$~\footnote{Thus, the entire analysis of the potential can be done assuming $\tilde\phi$ positive and straightforwardly extended to the negative semi-plane by reflection.}. In figure \ref{fig:slow_roll_monomial} is plotted the behavior with respect to $\tilde\phi$ for some arbitrary values of the exponent $n$. We can see from the plot of $\epsilon_V$ (left panel) that the smaller the value of $n$ the closer to zero is the value of $\tilde\phi$ at which there is the transition from the slow-roll regime to the end of inflation. The plot of $\eta_V$ instead shows the transition in the convexity of the potential when $n$ crosses $1$. Let us stress that observational data from the Planck collaboration favors concave potentials  \cite{Planck2018_const_infl}.

To determine the values of the scalar field at the critical moments of inflation, we can first use the condition end of inflation $\epsilon = 1$ to get $ \tilde\phi_{e} = \frac{n}{\sqrt{2}}$ and, using the equation \eqref{n_efolds}, we obtain the exit-mode and initial values of the field
\begin{equation}
\label{n_efolds2}
\tilde\phi_{(*,i)} =\sqrt{\frac{n^2+4nN_{(*,i)}}{2}},
\end{equation}
where we only consider the positive solution, for the sake of simplicity. The behavior of the negative solutions can be obtained by applying symmetry arguments, taking special care to the cases with $n<1$ (concave potentials). Now, using the equations \eqref{eq:spectral_index} - \eqref{eq:r}, we find the spectral index and the scalar-to-tensor ratio as follows
\begin{equation}
    n_s-1=-\frac{n(n+2)}{\tilde\phi_*^2},\quad \mbox{and}\quad r=\frac{8n^2}{\tilde\phi_*^2}.
\end{equation}

\begin{figure}[ht]
    \centering
    \includegraphics[scale=0.5]{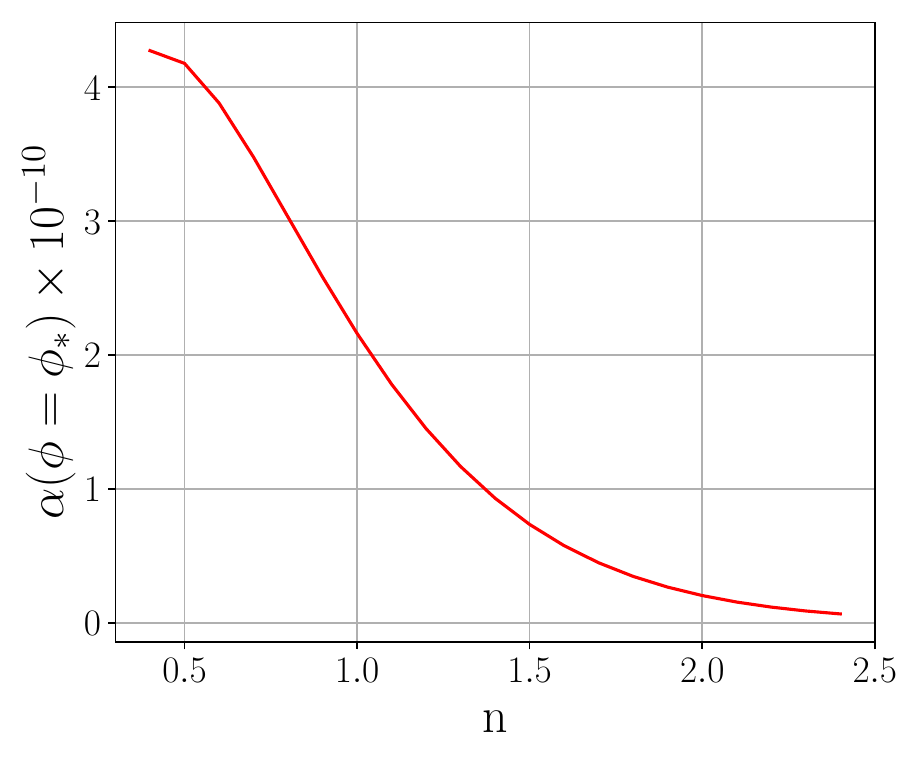}
    \caption{$\alpha$ as function of $n$ at horizon crossing mode for a monomial potential.}
    \label{fig:alpha}
\end{figure}

We use the expression of the power spectrum at the crossing horizon to calculate the amplitude of the potential in terms of $n$ through the equation \eqref{eq:P_r_phi}. Thus,
\begin{equation}
        \Delta^2_R(\phi)= \frac{\alpha\tilde\phi^{n+2}}{12 \pi^2 n^2}
\end{equation}
Setting $\Delta^2_R(\phi_*)=2.119 \times 10^{-9}$ \cite{Planck2018_grav_lens}, we find
\begin{equation}
\label{eq:ampl_mono_pot}
    \alpha=\frac{5,0193\times10^{-7}\times2^{\frac{n}{2}}n^2}{(4nN_*+n^2)^{\frac{n+2}{2}}}.
\end{equation}
The figure \ref{fig:alpha} shows the amplitude $\alpha$ as a decreasing function of $n$. Its variation is more sensitive around $n=1$, where the derivative is more negative and there is a change in the concavity.

\begin{figure}[!ht]
    \centering
    \includegraphics[scale=0.48]{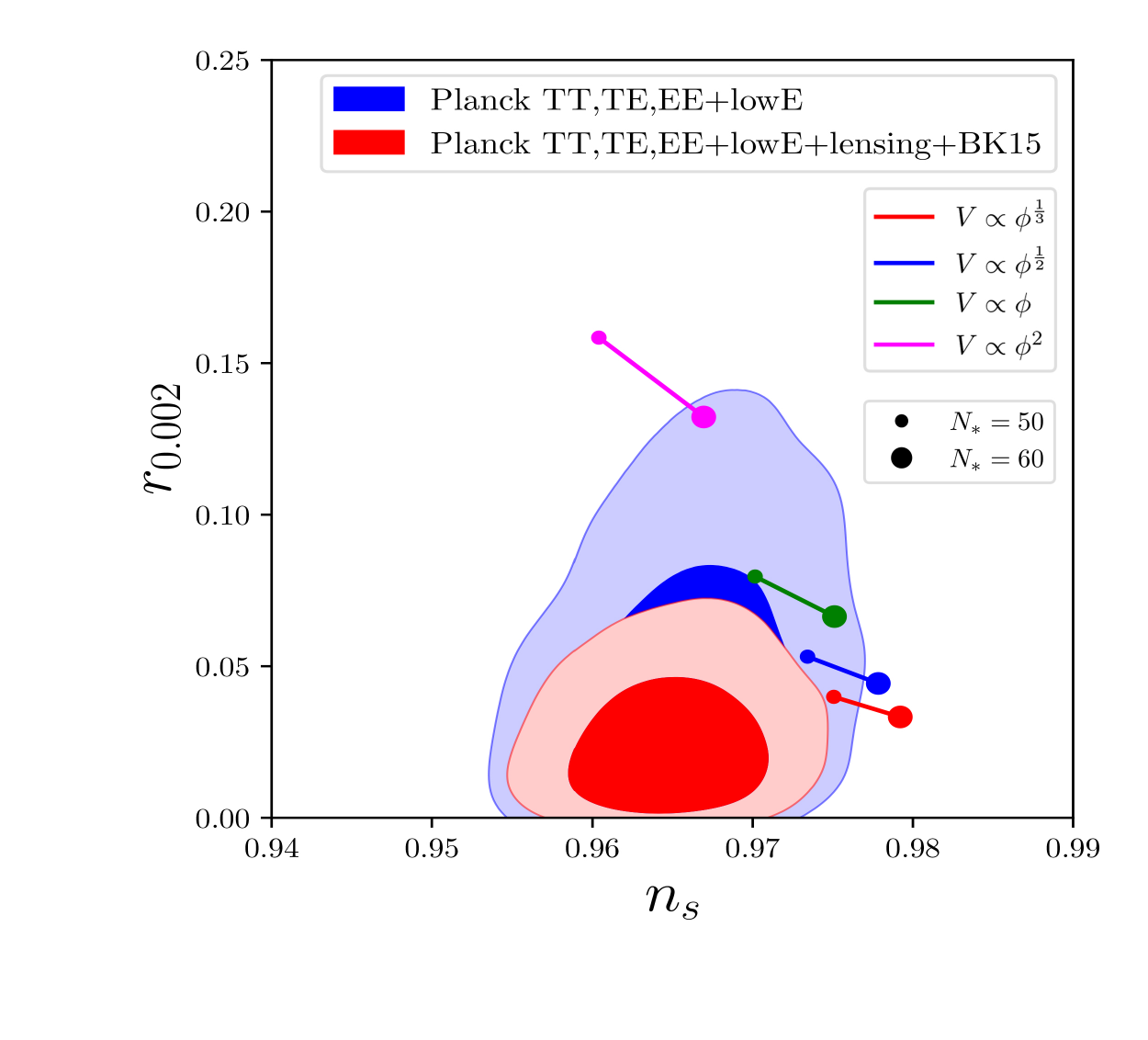}
        \caption{{$68\%$ and $95\%$ Confidence Levels for $n_s$ and $r$ parameters. The analysis refer to CMB Planck data 2018 (blue contours) and the extended dataset also including CMB lensing, BICEP2 and Keck Array experiments (red contours) \cite{Planck:2019nip,BICEP2:2018kqh}. We superimpose some case of monomial potentials, assuming $N_*\in[50,60]$.}}
    \label{fig:ns_r_monomial}
\end{figure}

At the same time, we explore the sensitivity of the inflationary parameter $n_s$ and $r$ with respect to the exponent $n$ in figure \ref{fig:ns_r_monomial}, for e-folds number in the interval $N_*\in[50,60]$. We note that concave and linear potentials with $N_*\sim50$ are preferred by Planck data 2018 \cite{Planck:2019nip}, while $n=2$ is allowed only for $N_*\sim60$, within $2\sigma$. Including tensorial observation by BICEP2 and Keck Array experiments \cite{BICEP2:2018kqh} the range of $r$ is narrowed, excluding all monomial potentials analyzed here at $2$ standard deviations in the range $N_*\in[50,60]$, requiring a lower number of e-folds to be consistent with the data at $95\%$ C.L.


\section{Binomial inflation}
\label{sec:binomial}

Let us now study the changes due to the introduction of a second polynomial term of higher order in the potential of the scalar field. Thus, we shall deal with
\begin{equation}
\label{eq:bi_potential}
    V (\phi) = \alpha M_{pl}^4\,\tilde{\phi}^n(1+\gamma\,\tilde{\phi}^{m-n}),
\end{equation}
where $m>n$, being both real number. With this choice, the first slow roll parameter becomes
\begin{equation}
\label{eq:epsilon_bi}
\epsilon_V=\frac{(m\gamma\,\tilde{\phi}^m + n\tilde{\phi}^n)^2}{2\,\tilde{\phi}(\gamma\,\tilde{\phi}^m + \tilde{\phi}^n)^2}.
\end{equation}
Again, the possible values of the scalar field at the end of inflation are obtained by setting $\epsilon_V\approx1$, yielding the following polynomial equation for $\tilde\phi$:
\begin{equation}
\label{eq:pol_epsilon}
2\gamma^2\,\tilde{\phi}^{2m+2} + 2\tilde{\phi}^{2n+2} -  m^2\gamma^2\,\tilde{\phi}^{2m} - n^2\tilde{\phi}^{2n}=0.
\end{equation}
However, it cannot be algebraically solved for arbitrary $m$ and $n$, and some extra hypothesis is needed for the exponents.

In order to proceed analytically with the study, we shall treat special cases for the exponents of the terms in the potential as in a Taylor expansion, combining their parity: for instance, in the case $(i)$ we consider $m=n+1$, which corresponds to a modification of the dominant term by other one with opposite parity; for the case $(ii)$, we choose $m=n+2$, which introduces a second term in the potential with the same parity of the leading one.

Any analytical inflationary model is expected to be approached this way, and the best-fit class of potentials could be perturbatively guest. This analysis allows us to understand how the next order in a Taylor expansion modifies the best fit of the monomial inflationary models. Then, we would have a hint of what the correct profile for the scalar field potential to look for is. The two cases of interest here are relevant because of their parity difference, emphasizing the role of symmetry in the potential terms for inflation parameters. The validity of the perturbative analysis is guaranteed as long as $\gamma$ is small. It should also be remarked that $\tilde\phi$ does not need to be small to validate the expansion. Actually, the potential must be valid for large $\tilde\phi$, when inflation begins. As $\tilde\phi$ diminishes, the inflation mechanism starts to leave the scene.

For the sake of completeness, the second-order slow-roll parameter is given by
\begin{equation}
\label{eq:eta_bi}
\eta_V=\frac{m(m-1)\gamma\,\tilde{\phi}^m + (n-1)n\tilde{\phi}^n}{\tilde{\phi}^2(\gamma\,\tilde{\phi}^m + \tilde{\phi}^n)},
\end{equation}
which is necessary to validate the slow-roll regime and, of course, in the calculation of the spectral index.

Finally, from the equation of the dimensionless power spectrum, we get the magnitude of the potential, as a function of $n$ and $\gamma$, as follows
\begin{equation}
\label{eq:alpha_bi}
    \alpha= \frac{12\pi^2\Delta^2_R\,(n + m\gamma\tilde\phi_*^{m-n-1})^2}{\tilde\phi_*^{n+2}(1+\gamma\tilde\phi_*^{m-n})^3}.
\end{equation}
Although it recovers the equation \eqref{eq:ampl_mono_pot} in the limit $\gamma\rightarrow0$, it is important to stress that $\phi_*$ would also have a contribution of order $\gamma$ is the case of binomial potential.

\subsection{Potentials with terms of opposite parity (m=n+1)}

As we said before, the study of binomial potentials cannot be done analytically for arbitrary exponents. For this reason, we shall study some particular cases of interest and try to infer qualitative properties of the inflation parameters that are invariant if other terms are added to the series.

We thus start by studying the relationship between terms of opposite parity in the scalar field potential, breaking the symmetry of the potential concerning reversion $\tilde\phi\rightarrow-\tilde\phi$. The simplest possible choice in which this holds and the calculations are manageable is considering the case $m=n+1$. Hence, the specific form of the potential of the equation \eqref{eq:bi_potential} is given by
\begin{equation}
\label{pot_bi_simp}
\frac{V (\tilde\phi)}{\alpha} =  M_{pl}^4\, \tilde{\phi}^n(1+\gamma\,\tilde{\phi}).
\end{equation}
Apart from $\tilde\phi=0$, the potential vanishes at $\tilde\phi=-1/\gamma$, which is irrelevant for the analysis if $\gamma$ is small (as we shall see, the interval of e-folds indicated by the data will lead to a limited range for $\tilde\phi$). A particular example here is the Starobinsky (S) model \cite{Starobinsky80}, whose potential is
\begin{equation}
V_{\text{S}}(\tilde\phi)=\Lambda^4\left(1-e^{\sqrt{\frac{2}{3}}\,\tilde\phi}\right)^2= \Lambda^4\left(\sqrt{\frac{2}{3}}\,\tilde\phi - \frac{\tilde\phi^2}{3}+\ldots\right).
\end{equation}

\begin{figure}[ht]
    \centering
    \includegraphics[scale=0.45]{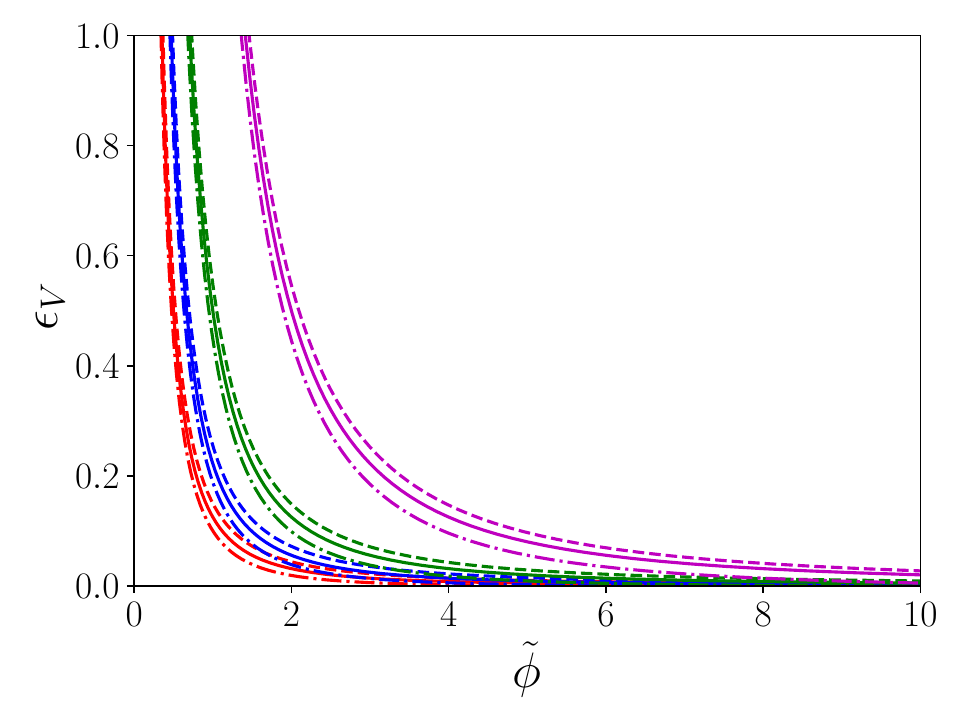}
    \includegraphics[scale=0.45]{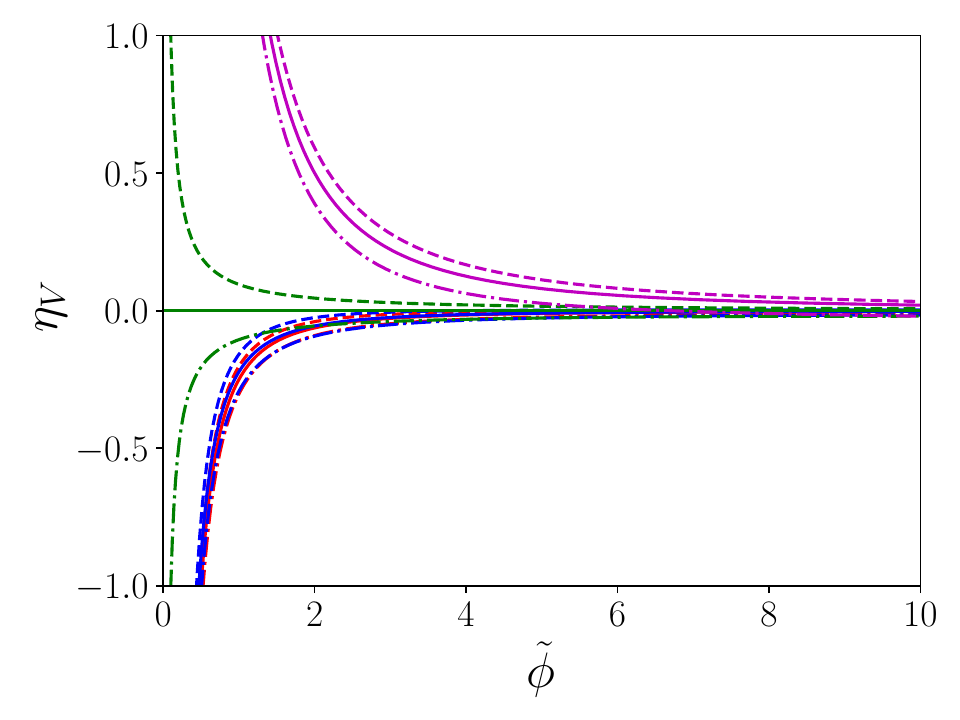}
    \caption{The slow-roll parameters $\epsilon_V$ and $\eta_V$ for the binomial potential with $m=n+1$. We consider $n=1/2$ (red), $n=2/3$ (blue), $n=1$ (green), and $n=2$ (magenta). Also, we use three values for $\gamma$: the dashed curves represent $\gamma=0.05$, the dash-dotted curves correspond to $\gamma=-0.05$ and the solid curves indicate $\gamma=0$.}
    \label{fig:e_d_binomial}
\end{figure}

In the case of potential of the equation \eqref{pot_bi_simp}, the slow-roll parameters, the equations \eqref{eq:epsilon_bi} - \eqref{eq:eta_bi}, take the form
\begin{equation}
\label{ep_bi_simp}
    \epsilon_V = \frac{[n + (n+1)\gamma\,\tilde{\phi}]^2}{2\,\tilde{\phi}^2(1 + \gamma\,\tilde{\phi})^2},
\end{equation}
and
\begin{equation}
\label{delta_bi_simp}
    \eta_V=\frac{n(n-1) + n(n+1)\gamma\,\tilde{\phi}}{\tilde{\phi}^2(1 + \gamma\,\tilde{\phi})}.
\end{equation}
Their behavior is depicted in figure \ref{fig:e_d_binomial} for some rational values of $n$ (convex and concave potentials) and small real values of $\gamma$. For each $n$, it is evident the moment characterizing the end of inflation, when the slow-roll parameters become the unit. The profile of the curves is quite similar to the monomial case, except for $n=1$ where the presence of a small $\gamma\neq0$ makes $\eta_V$ increase as $\tilde\phi$ goes to zero.

Under the condition $m=n+1$, the determination of $\phi_e$ comes from the roots of the equation \eqref{eq:pol_epsilon}, which is now a fourth-order polynomial given by
\begin{equation}
\label{polinomio2}
2 \gamma^2 \tilde\phi^4 + 4 \gamma \tilde\phi^3 + (2 - m^2 \gamma^2) \tilde\phi^2 - 2 nm \gamma \tilde\phi - n^2 = 0.
\end{equation}
The roots can be analytically obtained and expanded into a power series up to the first order in $\gamma$. But there are only two physical solutions for $\phi_e^{(b)}$ that are
\begin{equation}
\label{phi_e2}
    \tilde\phi_{e}^{(b)} = \pm \frac{n}{\sqrt{2}}+\frac{n}{2}\gamma.
\end{equation}
Note that the first-order correction in $\gamma$ is the same for both signs of $\tilde\phi_{e}^{(b)}$. The other two solutions are non-physical, diverging when $\gamma \rightarrow 0$.

We can now particularize the equation describing $\tilde\phi$ as a function of the number of e-folds, allowing us to calculate both the initial $\tilde\phi_i$ and the $\tilde\phi_*$ at the crossing horizon for the binomial case. From the equation \eqref{n_efolds}, we obtain
\begin{equation}
  N  = \int_{\tilde\phi_e^{(b)}}^{\tilde\phi} {\left(\frac{\tilde\phi(1+\gamma\tilde\phi)}{n+(1+n)\gamma\tilde\phi}\right)} d\tilde\phi \approx \int_{\tilde\phi_e}^{\tilde\phi} \left(\frac{\tilde\phi}{n}-\frac{\tilde\phi^2\gamma}{n^2}\right)d\tilde\phi,
   \label{eq:Nefold_bin_case_i}
\end{equation}
where the last expression was obtained assuming $\gamma$ small. By integrating this term by term and substituting $\tilde\phi_{e}^{(b)}$ explicitly, we can invert the resulting third-order polynomial equation to obtain $\tilde\phi(N)$. The outcome is
\begin{equation}
\tilde\phi_{(*,i)}^{(1)}=\pm\frac{\sqrt{n^2+4nN_{(*,i)}}}{\sqrt{2}}+\frac{(n+4N_{(*,i)})^2\pm 2n\sqrt{n^2+4nN_{(*,i)}}}{6(n+4N_{(*,i)})}\gamma,
\end{equation}
and
\begin{equation}
\tilde\phi_{(*,i)}^{(2)}=\pm\frac{\sqrt{n^2+4nN_{(*,i)}}}{\sqrt{2}} + \frac{(n + 4N_{(*,i)})^2 \mp 2n \sqrt{n^2+4nN_{(*,i)}}}{6(n+4N_{(*,i)})}\gamma,
\end{equation}
where $\tilde\phi_{(*,i)}^{(1)}$ is related to $\tilde\phi_e^{(+)}$, while $\tilde\phi_{(*,i)}^{(2)}$ is related to $\tilde\phi_e^{(-)}$. Henceforth, we shall deal only with the positive range of $\tilde\phi$, without loss of generality, supported by the monotonic behavior of the slow-roll parameters (see the figure \ref{fig:e_d_binomial}) and the smallness of $\gamma$, namely, the inflation mechanism ends before $\tilde\phi$ changes sign. Therefore, we shall use only the positive sign of $\tilde\phi_{(*,i)}^{(1)}$, neglecting $\tilde\phi_{(*,i)}^{(2)}$ which should give qualitatively the same. Again, we neglected the non-physical solutions due to divergences when $\gamma$ goes to zero.

\begin{figure}[ht]
    \centering
    \includegraphics[scale=0.45]{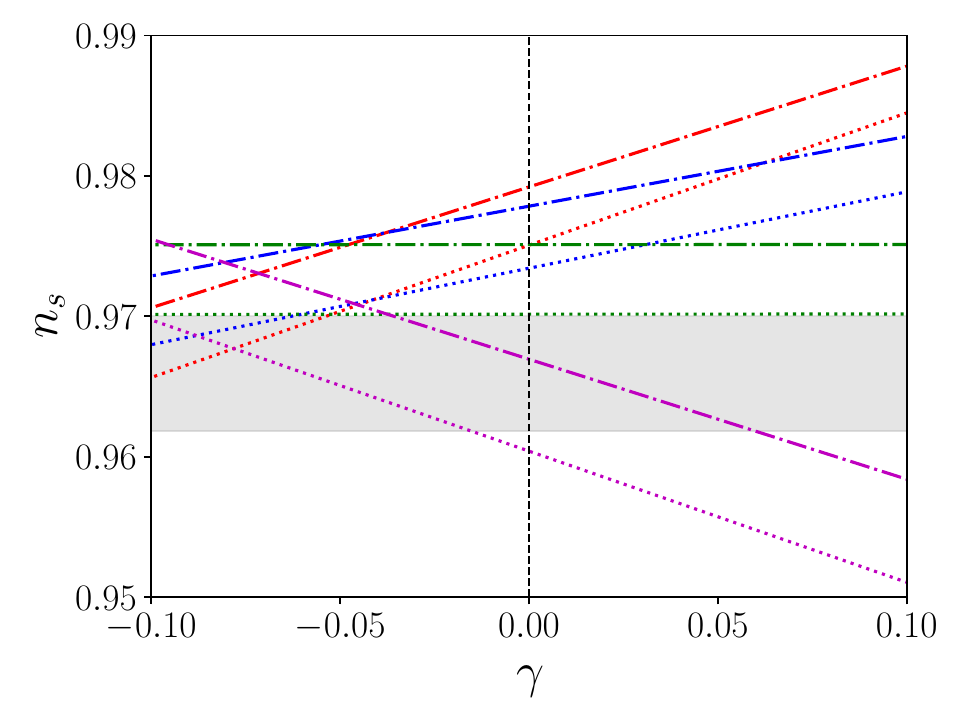}
    \includegraphics[scale=0.45]{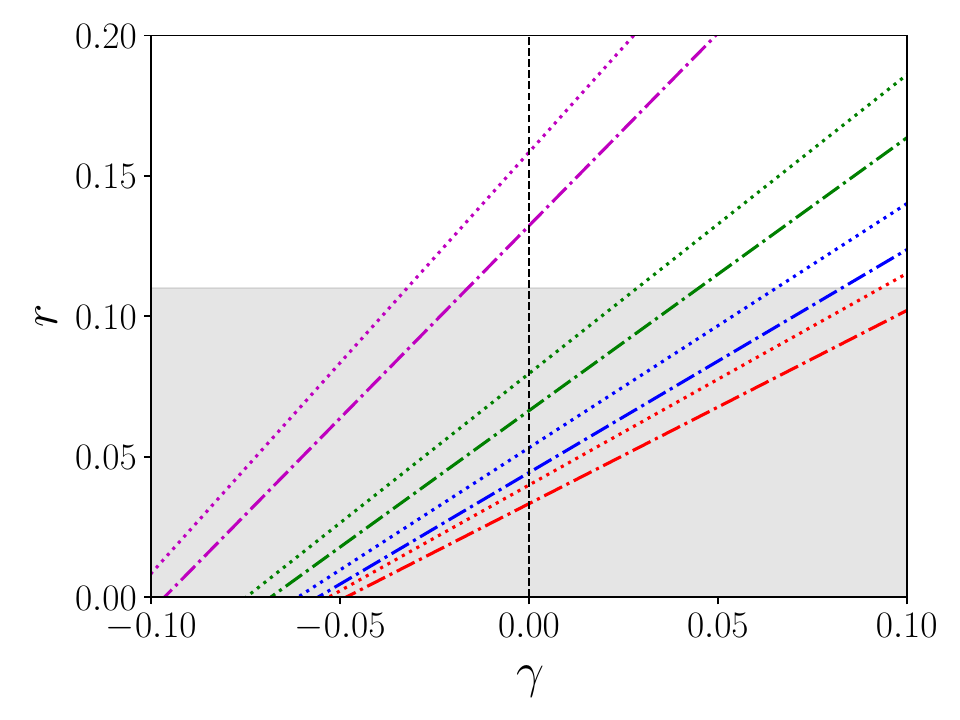}
    \caption{Scalar spectral index (left panel) and the tensor-to-scalar ratio (right panel) behaviour varying $\gamma$, assuming a positive sign of $\tilde\phi_{(*,i)}^{(1)}$. We consider $n=1/2$ (red), $n=2/3$ (blue), $n=1$ (green) e $n=2$ (magenta) with $N_*=50$ (dotted lines) and $N_*=60$ (dash-dotted lines). The gray region refers to the constrained values of Planck2018 data ($\Lambda$CDM+r model), i.e. $n_s=0.9659\pm 0.0041$ and $r<0.11$ \cite{Planck2018_const_infl}.}
    \label{fig:ns_r_2.1_binomial}
\end{figure}

The scalar spectral index and the tensor-to-scalar ratio, given by the equations \eqref{eq:spectral_index} - \eqref{eq:r}, respectively, can be written in their first-order form with respect to $\gamma$ as
\begin{equation}
    n_s-1=-\frac{2 (n+2)}{n + 4N_{*}} + \frac{4 \sqrt{2}\, \left[(2+n)n^{3/2} + (1-n)(n + 4N_{*})^{3/2}\right]}{3 \sqrt{n} (n + 4N_{*})^{2}}\,\gamma,
\end{equation}
and
\begin{equation}
    r = \frac{16 n}{n + 4N_{*}} - \frac{32 \sqrt{2} \sqrt{n} \left[n^{3/2} - (n + 4N_{*})^{3/2}\right]}{3 (n + 4N_{*})^{2}}\, \gamma.
\end{equation}
Figure \ref{fig:ns_r_2.1_binomial} depicts these parameters for different values of $n$ as a function of $\gamma$ and compared with the Planck data constraints in 1$\sigma$ \cite{Planck2018_const_infl}. From the left panel, the expected range of $n_s$ does not favor $\gamma>0$ for the exponents analyzed, except $n=2$ assuming $N_*=60$. For small $\gamma<0$, there is a good agreement only for $n=2$ with $N_*\in[50,60]$. The concave potentials ($n$=1/2 or 2/3) are also allowed, but they require a large negative $\gamma$ and $N_*$ around 50 to be in. The potential with a linear leading term ($n=1$) is marginally out for all $\gamma$. On the right panel, the upper bound of $r$ excludes $n=2$ for any $\gamma>0$ and $n=1$ for $\gamma>0.05$. The concave potentials are for a larger range of positive $\gamma$. For $\gamma<0$, the concave and linear potentials predict a very small scalar-to-tensor ratio for the whole interval $N_*\in[50,60]$, while it is needed $\gamma$ around $-0.05$ or less, to include $n=2$ in this set for all $N_*$ in the interval of interest. Of course, more restrictive constraints on these parameters would set stronger limits in the shape of the potentials \cite{Planck2018_const_infl}.

Finally, from the equation \eqref{eq:alpha_bi}, we can get the following expression for the magnitude of the dominant term of the potential using the dimensionless power spectrum amplitude at the pivot
\begin{equation}
\label{eq:alpha_bin_case_i}
    \alpha= \frac{12\pi^2\Delta^2_R\,[n + (n+1)\gamma]^2}{\tilde\phi_*^{n+2}(1+\gamma\tilde\phi_*)^3}.
\end{equation}
Note that the binomial case has a degeneracy in setting the correct amplitude of the potential through the power spectrum. Namely, we have only one equation to determine $\alpha$ and $\gamma$. A possibility to bypass this is to fix some values of $\gamma$ within the perturbative regime window and determine the corresponding $\alpha$ using the equation \eqref{eq:alpha_bin_case_i}. Thus, we can use a numerical analysis to construct the power spectrum for each case, and get the expected values of the cosmological parameters sensitive to the inflationary model.

In order to best constrain these values, we shall also consider the parameter $\sigma_8$, which gives the root mean square of the amplitude of matter perturbations smoothed over $8\,h^{-1}$Mpc, with $h$ as the Hubble constant in units of 100 km s$^{-1}$ Mpc$^{-1}$. It's value is $\sigma_8=0.811 \pm 0.006$ (68 \%, Planck TT, TE, EE + low E + lensing), according to Planck collaboration (2018) \cite{Planck2018_VIII}.  We can obtain this parameter value through numerical integration using a Boltzmann solver code. We choose to work with the \textit{Code for Anisotropies in the Microwave Background} (CAMB) \cite{Lewis:1999bs}, specifically utilizing its modified version ModeCode \cite{Mortonson:2010er, Easther:2011yq}  where the spectrum of CMB anisotropies is derived by numerically solving the equations governing inflationary modes. This includes addressing the Friedmann and Klein-Gordon equations, along with the Fourier components of the gauge-invariant quantity for a specific form of the single-field inflaton potential. In order to obtain our predictions, we use arbitrary values of cosmological parameters consistent with recent estimates,
fixing the physical baryon density, $\Omega_bh^2=0.0222$, the physical cold dark matter density, $\Omega_{cdm}h^2=0.1197$, the Hubble constant today, $H_0=67\, \mbox{km} \cdot \mbox{s}^{-1} \cdot \mbox{Mpc}^{-1}$, and the optical depth, $\tau= 0.08$.

\begin{figure}[!ht]
    \centering
    \includegraphics[scale=0.48]{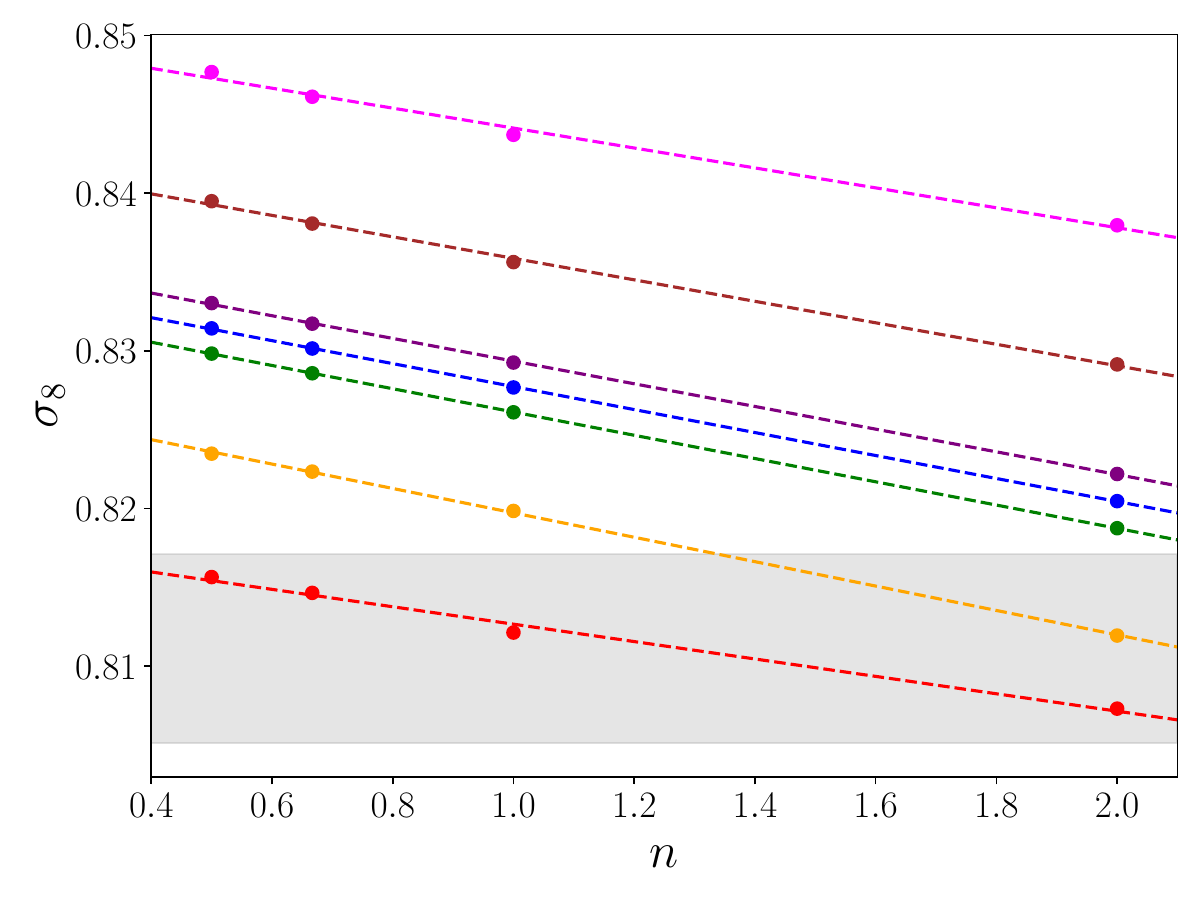}
    \caption{ The clustering parameter behavior with respect to the exponent $n$. The dots indicate the values of $\sigma_8$ for four different values of $n$ [$0.5$, $0.7$, $1$, $2$], while the dashed lines is the interpolation for each case with different $\gamma$'s. The intermediate line (blue) corresponds to $\gamma=0$. Below it we have $\gamma$ positive, starting with $1.0\times10^{-4}$ (green) and steps of $5.0\times10^{-4}$. The same is made for $\gamma<0$, indicated by the curves above the blue one, starting with $-1.0\times10^{-4}$ (purple) and steps of $-5.0\times10^{-4}$.
    The gray region refers to the constrained valued of Planck2018, $\sigma_8=0.811 \pm 0.006$ ($68\%$, Planck TT, TE, EE + low E + lensing).}
    \label{fig:sigma8_bi_1}
\end{figure}

The $\sigma_8$ parameter behavior is showed figure \ref{fig:sigma8_bi_1}, where the dashed lines correspond to different $\gamma$'s, while the dots indicate the values of $\sigma_8$ for the specific cases $n=$ 1/2, 2/3, 1 and 2. {The blue intermediate line represents $\gamma = 0$. Below this line, we have positive values of $\gamma$, starting at $1.0 \times 10^{-4}$ (in green) and increasing in steps of $5.0 \times 10^{-4}$. Above the blue line, we have negative values of $\gamma$, beginning at $-1.0 \times 10^{-4}$ (in purple) and decreasing in steps of $-5.0 \times 10^{-4}$. All exponents studied here achieves the confidence level with $\gamma$ around $5.0\times10^{-4}$.} On the other hand, $\gamma<0$ is significantly in tension with the expected value from the data, for any $n$ from 1/2 to 2 (at least in one standard deviation). It is important to emphasize that $\sigma_8$ is highly sensitive to $\gamma$. To make a meaningful comparison with the expected range of $\sigma_8$, we need to use $\gamma$ values that are two orders of magnitude smaller than those used for $n_s$ and $r$. Specifically, for $n_s$ and $r$, we use $\gamma\sim 10^{-2}$, while for $\sigma_8$, we set $\gamma\sim 10^{-4}$.

\subsection{Potentials with terms of same parity (m=n+2)}
We now consider the case with reversion symmetry in $\phi$ assuming the lowest difference between the exponents, namely, $m=n+2$. Clearly, this introduces a second term in the potential with the same parity as the leading term. Thus, the form of the potential \ref{eq:bi_potential} reduces to
\begin{equation}
\label{pot_bi_simp2}
\frac{V (\tilde\phi)}{\alpha} =  M_{pl}^4\,\tilde\phi^n(1+\gamma\tilde\phi^{2}).
\end{equation}
Again, there are particular examples that fit this potential at their lowest orders, which are the cases of Natural (N) inflation \cite{Freese1990,Adams1993}
\begin{equation}
V_{{\rm N}}(\phi)=\Lambda^4\left[1+\cos(\phi/f)\right] =\Lambda^4\left(2-\frac{\phi^2}{2f^2}+\ldots\right),
\end{equation}
and the Hilltop quadratic (Hq) model \cite{Boubekeur_2005}
\begin{equation}
V_{{\rm Hq}}(\phi)=\Lambda^4\left(1-\frac{\phi^2}{\mu^2}+\ldots\right).
\end{equation}

\begin{figure}[ht]
    \centering
    \includegraphics[scale=0.45]{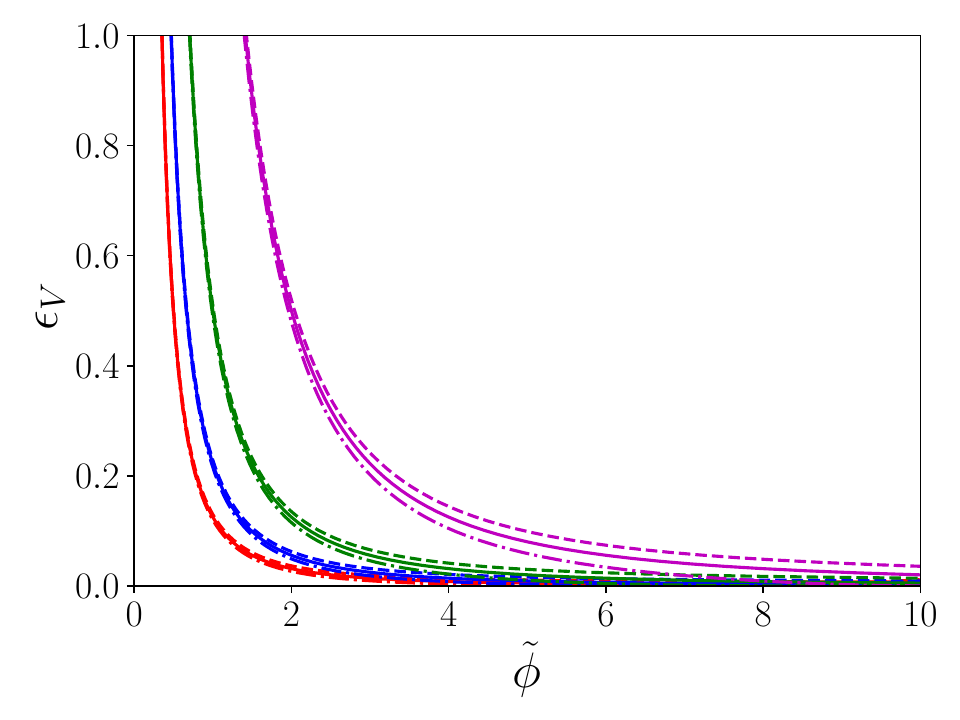}
    \includegraphics[scale=0.45]{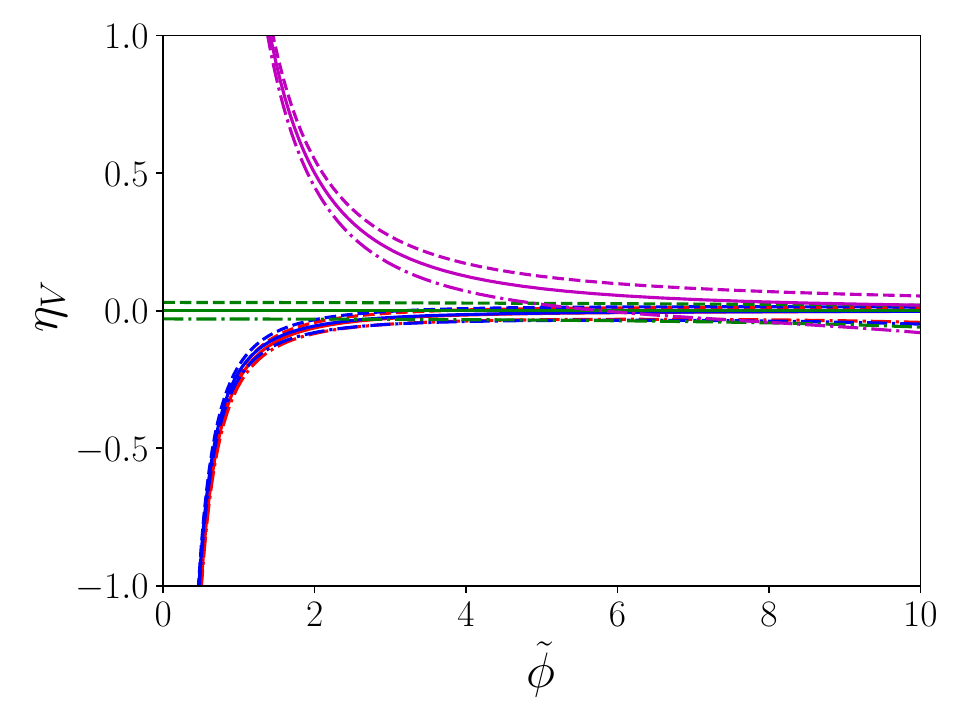}
    \caption{The slow-roll parameters $\epsilon_V$ and $\eta_V$ for the binomial potential with $m=n+2$. We consider $n=1/2$ (red), $n=2/3$ (blue), $n=1$ (green), and $n=2$ (magenta). Also, we use three values for $\gamma$: the dashed curves represent $\gamma=0.005$, the dash-dotted curves correspond to $\gamma=-0.005$ and the solid curves indicate $\gamma=0$. Note that the values chosen here are one order of magnitude less than the previous case as this case is more sensitive to $\gamma\neq0$.}
    \label{fig:e_d_binomial_2}
\end{figure}

By substituting the potential \ref{pot_bi_simp2} in the equation for the slow-roll parameters \ref{eq:epsilon_bi} and \ref{eq:eta_bi}, they are then written down as
\begin{equation}
    \epsilon_V = \frac{[n+(n+2)\gamma\tilde\phi^2]^2}{2\,\tilde\phi^2(1+\gamma\tilde\phi^2)^2},
\end{equation}
and
\begin{equation}
    \eta_V=\frac{(n-1)n+(n+1)(n+2)\gamma\tilde\phi^2}{\tilde\phi^2(1+\gamma\tilde\phi^2)}.
\end{equation}
In this case, their plots are given in the figure \ref{fig:e_d_binomial_2}. Similarly to the previous case, we used three values for $\gamma$, for different values of $n$. Again, analyzing the moment in which inflation ends, we see that $\eta_V$ does not increase its magnitude when $\tilde\phi$ goes to zero only for $n=1$. For $n\neq1$, the profile is qualitatively similar to the previous case. This is also a subtle behavior in the plots indicating that, for each $n$, $\gamma\neq0$ is more significant at the begin of inflation (for large $\tilde\phi$'s) since the relative distances of the curves diminish as inflation ends ($\epsilon\rightarrow 1$).

To find $\tilde\phi_e$, we substitute the condition $m=n+2$ into the equation \ref{eq:pol_epsilon}, obtaining
\begin{equation}
2\gamma^2\tilde\phi^6+[4-\gamma(n+2)^2]\gamma\tilde\phi^4+[2-2\gamma(n^2+2n)]\tilde\phi^2-n^2=0.
\end{equation}
This cubic polynomial in $\tilde\phi^2$ can still be solved analytically and the solutions expanded into a power series in $\gamma$ up to the first order. From the six possible values that $\phi_e$ admits, only
\begin{equation}
\label{phi_e_case_ii}
    \phi_e^{(b)}=\pm\frac{n}{\sqrt{2}}\pm\frac{n^2}{\sqrt{2}}\,\gamma
\end{equation}
are physically acceptable. The other four solutions diverge when $\gamma \rightarrow 0$, not recovering the monomial case perturbatively.

Now, we can find the equation relating $\tilde\phi$ and the number of e-folds, determining the values of the scalar field at the beginning of inflation and at the horizon crossing instant. Particularizing the equation \ref{n_efolds} to this case, we obtain
\begin{equation}
  N  = \int_{\tilde\phi_e}^{\tilde\phi} \frac{\tilde\phi(1+\gamma\tilde\phi^2)}{n+(n+2)\gamma\tilde\phi^2} d\tilde\phi\approx\int_{\tilde\phi_e}^{\tilde\phi} \left(\frac{\tilde\phi}{n}-\frac{2\tilde\phi^3\gamma}{n^2}\right)d\tilde\phi,
   \label{eq:Nefold_particular}
\end{equation}
where $\gamma$ is considered small in the last equation. Again, integrating term by term and using only the positive domain for $\tilde\phi$ during the inflation, as indicated by the slow-roll parameters, we take the positive sign of $\tilde\phi_e$, given by the equation \ref{phi_e_case_ii}, to get the preliminary equation
\begin{equation}
    N=\frac{n\left[\tilde\phi^2-\frac{1}{4}n^2(\sqrt{2}-\gamma)^2\right] + \left[\frac{1}{16}n^4(\sqrt{2}-\gamma)^4-\tilde\phi^4\right]\,\gamma}{2n^2}.
\end{equation}
Thus, by solving this biquadratic equation in $\tilde\phi(N)$ and series expanding in $\gamma$, we encounter
\begin{equation}
  \tilde\phi_{*,i}=\frac{\sqrt{n^2+4nN_{*,i}}}{\sqrt{2}}+\frac{\sqrt{n}(-n+2\sqrt{2}nN_{*,i}+4\sqrt{2}N_{*,i}^2)}{2(\sqrt{n+4N_{*,i}})}\,\gamma,
\end{equation}
where all the other possible solutions were neglected by the same physical reasons as in the previous case.

\begin{figure}[ht]
    \centering
     \includegraphics[scale=0.45]{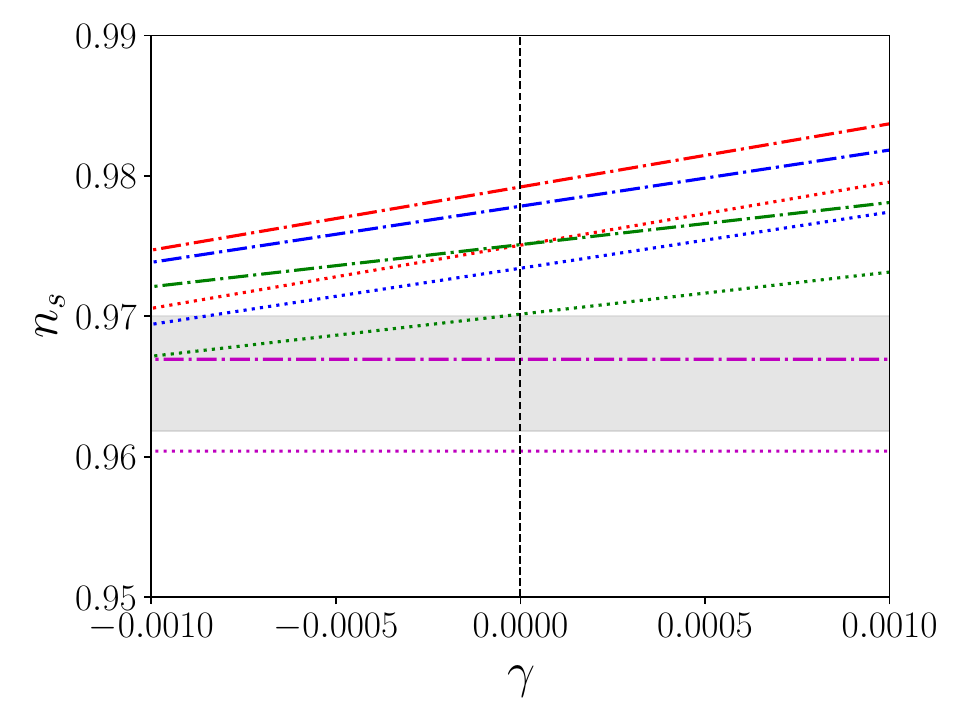}
      \includegraphics[scale=0.45]{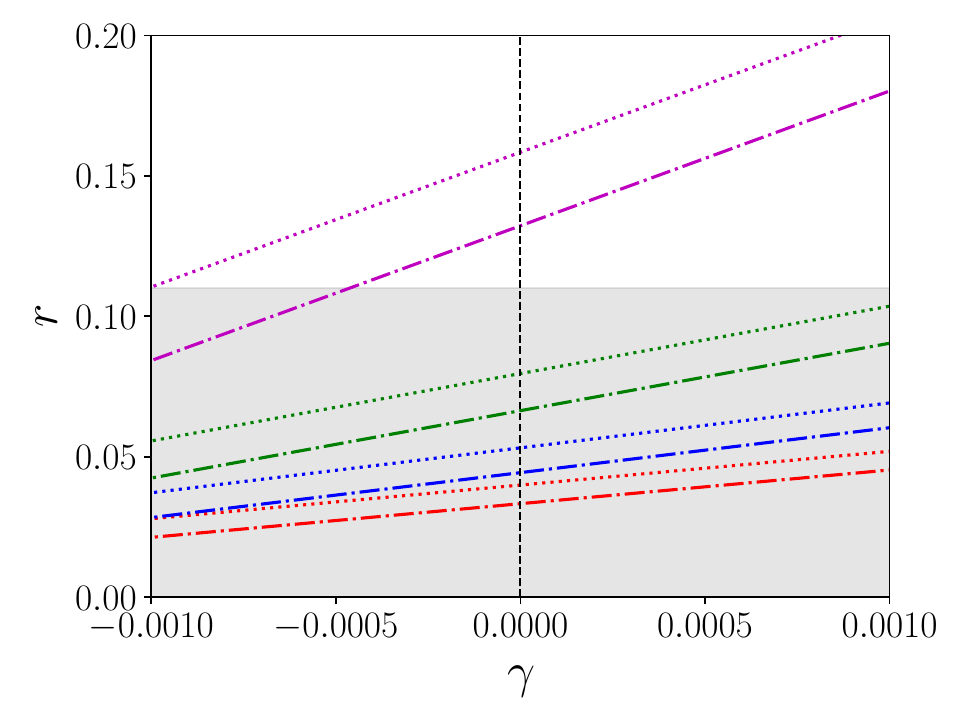}
   \caption{ Scalar spectral index (left panel) and the tensor-to-scalar ratio (right panel) behaviour varying $\gamma$. We consider $n=1/2$ (red), $n=2/3$ (blue), $n=1$ (green) e $n=2$ (magenta) with $N_*=50$ (dotted lines) and $N_*=60$ (dash-dotted lines). The gray region refers to the constrained values of Planck2018 data ($\Lambda$CDM+r model), i.e. $n_s=0.9659\pm 0.0041$ and $r<0.11$ \cite{Planck2018_const_infl}.}
    \label{fig:ns_r_binomial2}
\end{figure}

The scalar spectral index and the tensor-to-scalar ratio, given respectively by the equations \ref{eq:spectral_index} and \ref{eq:r}, are written now as follows
\begin{equation}
\label{eq:n_s_bin_case_2}
    n_s-1=-\frac{2 (2 + n)}{n + 4N_{*}} + \left(6 - 3n - \frac{n (2 + n) (2 \sqrt{2} + n)}{(n + 4N_{*})^2}\right) \gamma,
\end{equation}
and
\begin{equation}
    r=\frac{16 n}{n + 4N_{*}} + 8 n \left(3 + \frac{n (2 \sqrt{2} + n)}{(n + 4N_{*})^2}\right) \gamma.
\end{equation}
The figure \ref{fig:ns_r_binomial2} depicts their behavior as a function of $\gamma$, for different values of $n$ and $N_*$. First, we see that $\gamma\neq0$ does not alter the predictions of $n=2$ for $n_s$, which is in agreement with the parameters mean values by the way, since the first two terms within the parenthesis in the equation \eqref{eq:n_s_bin_case_2} cancel each other out and $N_*$ suppresses the remaining one. It does not happen for $n\neq2$, and the concave potentials are only marginally in agreement with the 1$\sigma$ range of $n_s$ (the most favored exponent is $n=1/2$, with the biggest inclination for $n_S(\gamma)$. For $n=1$, there is an intersection with the CMB constraint for $N_*<60$. On the other hand, $\gamma>0$ is excluded for all the potentials under study with $n<2$. Second, the behavior of the scalar-to-tensor ratio leads to similar qualitative conclusions compared to $n_s$, but now the $n=2$ case is marginally in agreement with the constraints on $r$. All the other values of $n$ are allowed for any $\gamma<10^{-3}$.

\begin{figure}[!ht]
    \centering
    \includegraphics[scale=0.48]{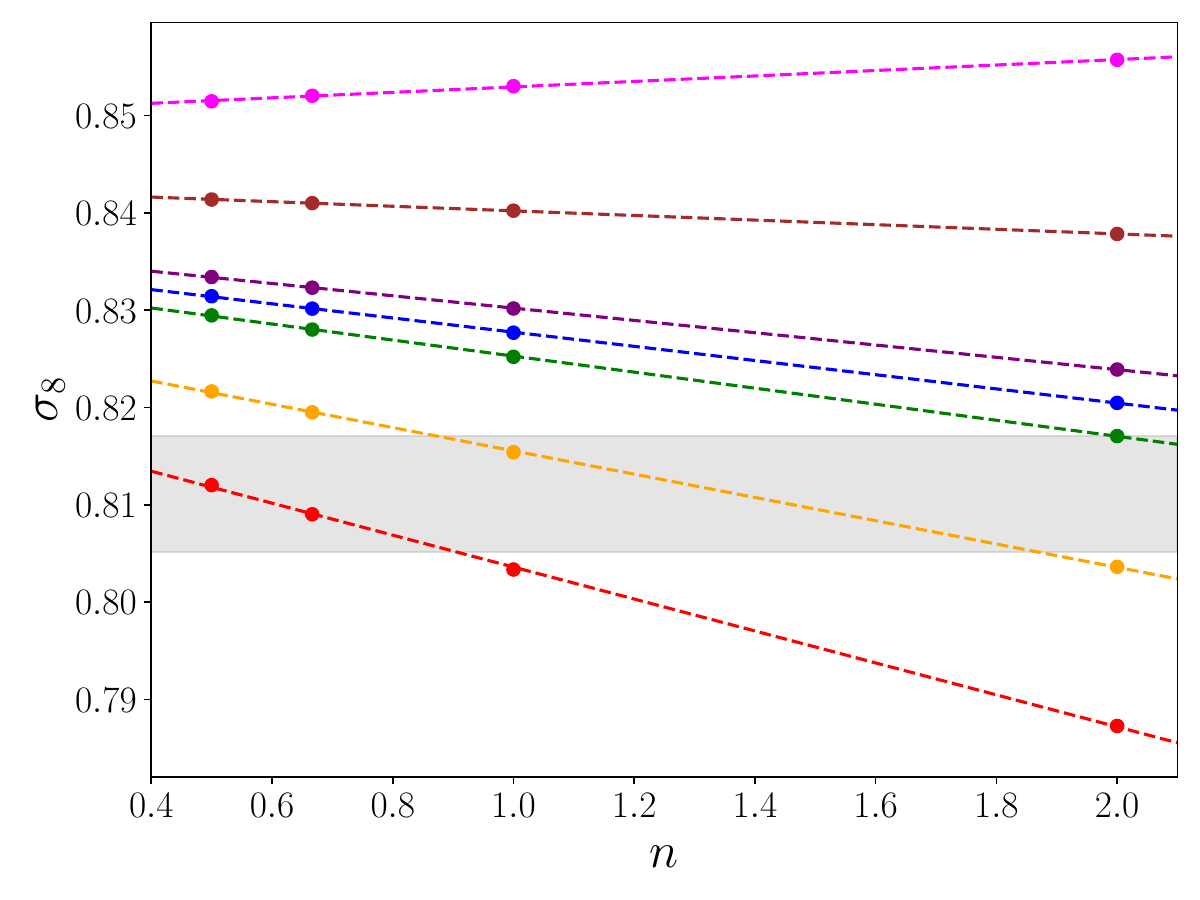}
    \caption{The clustering parameter behavior with respect to the exponent $n$. The dots indicate the values of $\sigma_8$ for four different values of $n$ [$0.5$, $0.7$, $1$, $2$], while the dashed lines is the interpolation for each case with different $\gamma$'s. The intermediate line (blue) corresponds to $\gamma=0$. Below it we have $\gamma$ positive, starting with $1.0\times10^{-5}$ (green) and steps of $5.0\times10^{-5}$. The same is made for $\gamma<0$, indicated by the curves above the blue one, starting with $-1.0\times10^{-5}$ (purple) and steps of $-5.0\times10^{-5}$.
    The gray region refers to the constrained valued of Planck2018, $\sigma_8=0.811 \pm 0.006$ ($68\%$, Planck TT, TE, EE + low E + lensing).}
    \label{fig:gamma_n_bi2}
\end{figure}

Finally, we calculate the amplitudes of the potential using the dimensionless power spectrum using the equation \ref{eq:P_r_phi}, yielding
\begin{equation}
\label{eq:alpha_bin_case_ii}
    \alpha= \frac{12\pi^2\Delta^2_R\,[n + (n+2)\gamma\phi]^2}{\tilde\phi_*^{n+2}(1+\gamma\tilde\phi_*^2)^3},
\end{equation}
which is degenerated of course, as we have one equation for the $\alpha$ and $\gamma$ variables. To bypass this, we again set a range of values for $\gamma$ and determine the corresponding $\alpha$ using the equation \ref{eq:alpha_bin_case_ii}. We then run the numerical codes to get the cosmological parameters predicted by each potential and use $\sigma_8$ to confront them. The results are presented in the figure \ref{fig:gamma_n_bi2}. Again, the blue dashed line corresponds to $\gamma=0$, while above it we have $\gamma<0$, and below it is $\gamma>0$. The difference in $\gamma$ from one curve to the other is now $5.0\times10^{-5}$ starting with $1.0\times10^{-5}$ in modulo. {For $\gamma<0$, all exponents predict $\sigma_8$ out of the window allowed. On the other hand, for $\gamma>0$, all exponents have an interval for $\gamma$ in agreement with CMB data. In particular, the concave potentials require $\gamma\sim10^{-4}$, while the linear and convex potentials require only $\gamma\sim10^{-5}$. This means that concave potentials tend to demand values of $\gamma$ greater than convex ones. It should also be noted that this is happening one order of magnitude less than the previous case in $\gamma$, since the inflationary parameters are even more sensitive under the condition $m=n+2$.}

\section{Concluding remarks}
\label{sec:results}

In this paper we have studied the sensitivity of the binomial inflationary potential spectral index $n_s$, scalar-to-tensor ratio $r$, and $\sigma_8$ parameters with respect to perturbations of a monomial potential in an inflationary model driven by a single scalar field. First, we briefly review the inflationary physics, specifically the slow-roll regime for a monomial potential. We demonstrate that the CMB+BK15 data impose strong constraints on the exponent's value for various numbers of e-folds. Next, we introduce a second term and explore binomial potentials. After some manipulation, it becomes clear that an analytical analysis for arbitrary exponents is not feasible. Therefore, we focus on two specific cases where the relationship between the exponents of the two terms is determined by their parity (even or odd). We investigate scenarios in which the terms have either opposite or the same parity, both of which are well-known in the literature. Since the slow-roll regime requires a scalar field to move slowly toward the potential's minimum, it is reasonable to include low-order terms in the potential's Taylor expansion. This leads us to expect relationships such as $m=n+1$ in one case and $m=n+2$ in the other, applicable to a broad class of potentials that guide the inflationary mechanism. Finally, we aim to identify qualitative properties of the inflationary parameters that might remain invariant regardless of the addition of other terms in the series.

The task to pass all criteria is much more complicated than it seems as each parameter has its preferred tendency concerning perturbations. For instance, the first case studied, namely, a binomial potential with exponents satisfying $m=n+1$, shows that in order to match the CMB observations on $n_s$ and $r$, negative amplitudes are constrained while indications on $\sigma_8$ behave exactly the opposite. Something similar occurs for the second case ($m=n+2$), but the discrepancy is more pronounced since we work with perturbations one order of magnitude smaller than the first case.

{In the end, we conclude that: (i) concave potentials are good in predicting low $r$, but the predictions for $n_s$ and $\sigma_8$ are contradiction; the former requires positive $\gamma$ while the latter demands $\gamma<0$; (ii) the linear potential behaves the same, but the conflict in the predictions for $n_s$ and $\sigma_8$ can be solved by diminishing the number of e-folds with perturbations of opposite parity; (iii) the convex potential (n=2) presents the best scenario for $n_s$ and $\sigma_8$ at small perturbations, where the only issue lays is the prediction of $r$ larger than the upper bound suggested by CMB data. In principle, this could be alleviated by increasing the number of e-folds, mainly for the case of perturbations with the same parity.}

If we attempt to go beyond the perturbative regime by introducing an extra term in the scalar field potential, the analytical approaches would be limited and insufficient, as several equations would not have algebraic solutions. Thus, a numerical analysis would be necessary. On the other hand, there would be no more fundamental justification for the insertion of such a term without a more fundamental physics explaining the shape of the potential. Therefore, it would be interesting to investigate in the future more sophisticated inflationary models guided by scalar fields, with the first analysis being the perturbative consistency of their parameters, as presented here.

\acknowledgments
The authors are thankful for the support of CNPq (E.B. grant N.\ 305217/2022-4), CAPES (M.E.S.A. grant N.\ 671304/2022-0) and Fapemig (F.A.F. grant BIC 7788).
MB is supported by the Istituto Nazionale di Fisica Nucleare (INFN), Sezione di Napoli, iniziativa specifica QGSKY.

\bibliography{ref}

\end{document}